\documentclass[twocolumn]{aastex701}
\usepackage{graphicx}
\usepackage{txfonts}

\usepackage{siunitx}
\usepackage{comment}
\usepackage{textcomp}
\usepackage[utf8]{inputenc}
\usepackage{multirow}
\usepackage{url}

\begin{document}

\title[BL Lac host galaxies characterisation]{BL Lac host galaxies: how to systematically characterise them in
optical-NIR spectroscopy}



\author[]{Gaia Delucchi}
\affiliation{Dipartimento di Fisica, Università di Genova, 
              Via Dodecaneso 33, 16146, Genova, Italy}
\affiliation{INFN - Sezione di Genova, Via Dodecaneso 33, 16146, Genova, Italy}
\affiliation{INAF - Osservatorio Astronomico di Brera, Via Bianchi 46, I-23807 Merate, Italy}
\email[show]{gaia.delucchi@ge.infn.it}  

\author[]{Tullia Sbarrato}
\affiliation{INAF - Osservatorio Astronomico di Brera, Via Bianchi 46, I-23807 Merate, Italy}
\email[]{tullia.sbarrato@inaf.it}  

\author[]{Giorgio Calderone}
\affiliation{INAF-Osservatorio Astronomico di Trieste, Via G.B. Tiepolo, 11 I-34143 Trieste, Italy}
\email[]{giorgio.calderone@inaf.it}  

\author[]{Chiara Righi}
\affiliation{INAF - Osservatorio Astronomico di Brera, Via Bianchi 46, I-23807 Merate, Italy}
\email[]{chiara.righi@inaf.it}  

\author[]{Silvano Tosi}
\affiliation{Dipartimento di Fisica, Università di Genova, 
              Via Dodecaneso 33, 16146, Genova, Italy}
\affiliation{INFN - Sezione di Genova, Via Dodecaneso 33, 16146, Genova, Italy}
\email[]{Silvano.Tosi@ge.infn.it}  

\author[]{Boris Sbarufatti}
\affiliation{INAF - Osservatorio Astronomico di Brera, Via Bianchi 46, I-23807 Merate, Italy}
\email[]{boris.sbarufatti@inaf.it}  


 
\begin{abstract}
  Host galaxies of Active Galactic Nuclei give crucial information on the interaction between accreting Supermassive Black Holes and their surroundings, and on their common evolution. 
  Their study in the case of aligned jetted AGN - BL Lacertae objects in particular - is complicated by the non-thermal jet component, whose bright and multi-frequency emission easily dominates over the whole electromagnetic spectrum. 
  BL Lac host galaxies have thus been sparsely studied, and their elliptical nature is currently a hypothesis supported by few observations. 
  With the broad aim of a systematic analysis of these sources, and in light of the many optical and NIR spectroscopic facilities that are now available, we implement an easily applicable method to determine whether a BL Lac is hosted in an elliptical or spiral galaxy. 
  Building on the only systematic study currently available on BL Lac hosts, we worked on a sample of realistic BL Lac synthetic spectra. 
  We analysed them and characterized their statistics using QSFit, a publicly available spectroscopy software.
  If BL Lac host galaxies were both elliptical and spiral, our method would be able to discriminate between them, provided that BL Lac jets are fainter than $L_\gamma\sim10^{46}$erg/s. 
  Just two runs of QSFit for each BL Lac spectrum would return a single parameter, that would allow for a first broad distinction between the two classes. 
  We finally discuss the two galaxy types that introduce some uncertainty in their classification, that might lead to possible classification biases. 
\end{abstract}


%

\section{Introduction}

Accretion of matter on Supermassive Black Holes (SMBHs) at the center of galaxies gives rise to a wide variety of energetic multiwavelength phenomena, which are collectively referred to as Active Galactic Nuclei \cite[AGN,][]{urry95,padovani17}.
AGN have a wide range of observational features, ultimately ascribable to a few intrinsic factors: their accretion rate, the viewing angle under which we observe them, their obscuration, and the presence of a relativistic jet.
The latter has been historically associated with a radio luminosity in stark excess with respect to the optical emission: radio-loud sources are expected to host relativistic jets, responsible for a radio emission so bright that dominates over the optical light radiated by the nucleus \citep{fanaroffriley74}.
This radio-loudness based definition is now being abandoned in favor of a more fundamental jetted vs.\ non-jetted AGN classification \citep{padovani17b}.

Among the jetted sources, blazars are a class defined only by their orientation: they are classified as AGN with their relativistic jets oriented close to our line-of-sight. 
Since the jet is thought to be aligned with the SMBH angular momentum, and hence likely perpendicular to the accretion flow plane, their orientation minimizes the absorption of their nuclear emission by the host dust in its immediate surrounding. 
Blazars offer a privileged point of view on nuclear emission, allowing observations of all the possible accretion regimes of jetted sources. 
Blazars are usually further classified into Flat Spectrum Radio Quasars (FSRQ) or BL Lacertae (BL Lac) objects, depending on their features in the optical-UV wavelength range: FSRQs are characterized by strong emission lines and a bright, blue continuum, while BL Lacs show no or very weak broad emission lines, which is thought to be a proof of their low accretion rate. 
This is confirmed by other tracers of accretion luminosities, such as intrinsic line brightness and relativistic jet power \citep{ghisellini14}: BL Lacs are likely radiatively inefficient accreting SMBHs, with their jet pointed close to our line of sight.
Their orientation and intrinsic power have made them optimal emitters of high-energy radiation, up to the TeV range. 

The emission from blazars relativistic jets, in fact, dominates across almost the whole electromagnetic spectrum, from radio frequencies to $\gamma$-rays (TeV included).
Their Spectral Energy Distribution (SED) is characterized by two prominent humps. 
The lower frequency hump can peak from the radio-sub-mm regime to the X-rays, depending on the nature of the source (typically FSRQs at lower frequencies, BL Lacs at larger ones), and is generally associated to synchrotron emission. 
The higher frequency hump peaks instead from the MeV-GeV energies (FSRQs) up to the TeV regime (extreme BL Lacs).
Many models describing different jet composition and emission channels has been proposed in the last $\sim30$ years, debating on the available energetic budget and emitting features of the competing approaches \citep[see e.g.][the last two for reviews]{maraschi92,dermer95,aharonian00,muecke03,ghisellini05,ghisellini09,boettcher19,ghisellini19}. 
While a general consensus is yet to be reached, the broad emission profile is agreed upon, and can be simply described with a combination of smoothly broken power-laws \citep{ghisellini17}.
This allows for a rough estimate of the radiative output of blazars relativistic jets, even if lacking a solid physical description. 
Beyond this general agreement, many of the details of the blazar class are still open questions, with some of them extending to the wider AGN family. 
Their evolutionary track is one of them: it is not yet clear whether different AGN classes correspond to different evolutionary stages. 
A complete view on the AGN host galaxies and environments, along with a deeper knowledge of their central engines, could be very helpful in this perspective.

Indeed a long standing open problem regarding AGN is the relation between the central SMBH and its host galaxy:
correlations between the respective masses, feedback tracers, and outflows have been thoroughly studied for more than 25 years \citep[see][for a review]{kormendy13}, most of them suggesting a connection. 
However there is currently no consensus on how they actually influence each other, whether their common evolution is dominated by AGN radiation or by the host galaxies gravitational potential.
A hint on the possible evolutionary patterns can be derived from AGN host galaxies, their type, their mass and their stellar population age. 

AGN can be hosted by all kinds of galaxies, while jetted AGN seem to inhabit mainly giant elliptical galaxies. 
Blazars have been historically associated only with elliptical host galaxies, but the identification of their host galaxies has also been extremely challenging, since the bright non-thermal flux in the NIR-optical-UV wavelength range often dominates over the expected host galaxies emission. 
Discriminating host and jet signatures is a challenging task, often limited by the complete absence of absorption or emission features, that does not allow for a spectroscopic redshift derivation and thus for detailed analysis and spectral decomposition. 
A careful host identification has been historically tackled with two approaches: (i) accurate imaging in the visible range, that allows for jet and host decomposition \citep[e.g.\ {\it Hubble} Space Telescope imaging][]{urry00,Sbarufatti_2005}, or (ii) analysis of reasonable signal-to-noise ratio (S/N) spectra that include absorption features due to the host, or significant host-to-jet dominance \citep{shaw13,pita14,goldoni21,kasai23}.

BL Lacs are thus found to be hosted in ``standard" elliptical galaxies, with rather narrow absolute magnitude distributions ($\pm0.5$mag), peaked between $M_R=-22.9$ \citep{Sbarufatti_2005} and $-22.5$ \citep{shaw13}. 
Notably, \citet{pita14} derive host galaxies as bright as $M_R=-24$, with an average brightness compatible with results by \citet{Sbarufatti_2005}. 
More recent results are still in line with the average $M_R$, but find wider dispersions \citep[up to 1.0 in][]{kasai23}.

The nature of the only spiral candidate found to date \cite[PKS 1413+135,][]{perlman02} is currently under debate. 
The source is proposed to be an intriguing chance alignment between a foreground spiral galaxy and a background lensed BL Lac \citep{readhead21}, but further investigation is needed to settle the matter.

The narrow magnitude distributions found in literature for BL Lac host galaxies allowed many groups to use them as standard candles in imaging studies, and thus derive photometric redshifts in case of featureless spectra \citep[e.g.][]{nilsson24}, always assuming them to be elliptical.
No systematic study on possible biases on the galaxy type has been performed, up to date.
With this work, we aim at exploring this possibility. 

The spectroscopic facilities that have been launched and built in the recent years (e.g.\ JWST, {\it Euclid}, LSST) give the unprecedented opportunity of a better view on blazars host galaxies. 
In the following, we show how we can get ready to tackle the study of BL Lacs host galaxies with a systematic approach. 
We aim at producing an easy-to-use algorithm and a set of statistic parameters that will allow us to discriminate whether a BL Lac is hosted by a spiral or an elliptical galaxy, opening the way to further in-depth studies on their age, stellar population and overall features.  
The readiness-to-use of our approach relies only on a single UV-optical-NIR spectrum and an open source software \cite[QSFit,][]{calderone17}.
We produced a set of synthetic spectra, generated by considering both a range of BL Lac luminosities and a set of different galaxy templates (Section \ref{section:synth_spec}). 
We then set up a simplified spectroscopic analysis using the open source software QSFit (Section \ref{section:3}) to obtain statistical characterisation for our simulated sample.
Finally, in Section \ref{section:4} we introduce the conditions to identify a BL Lac host galaxy, and ultimately an easy systematic approach only based on two QSFit runs.


\section{Synthetic BL Lac + host spectra}
\label{section:synth_spec}

The most recent  investigation specifically oriented at identifying the host galaxy population of BL Lacs dates back to about 20 years ago with \cite{Sbarufatti_2005}. The small data sample they used supported the prevalent hypothesis that only elliptical galaxies can host BL Lacs. In this context, we aimed to create a reliable method to identify and classify the hosts of these sources to confirm or debunk this theory. 
In order to test and calibrate the method, we first produced a set of  synthetic spectra covering a large number of possible combinations of host galaxy types, BL Lacs and galaxies luminosities, located at different redshifts $z$. 

\begin{figure}[h!]
    \includegraphics[width=9cm]{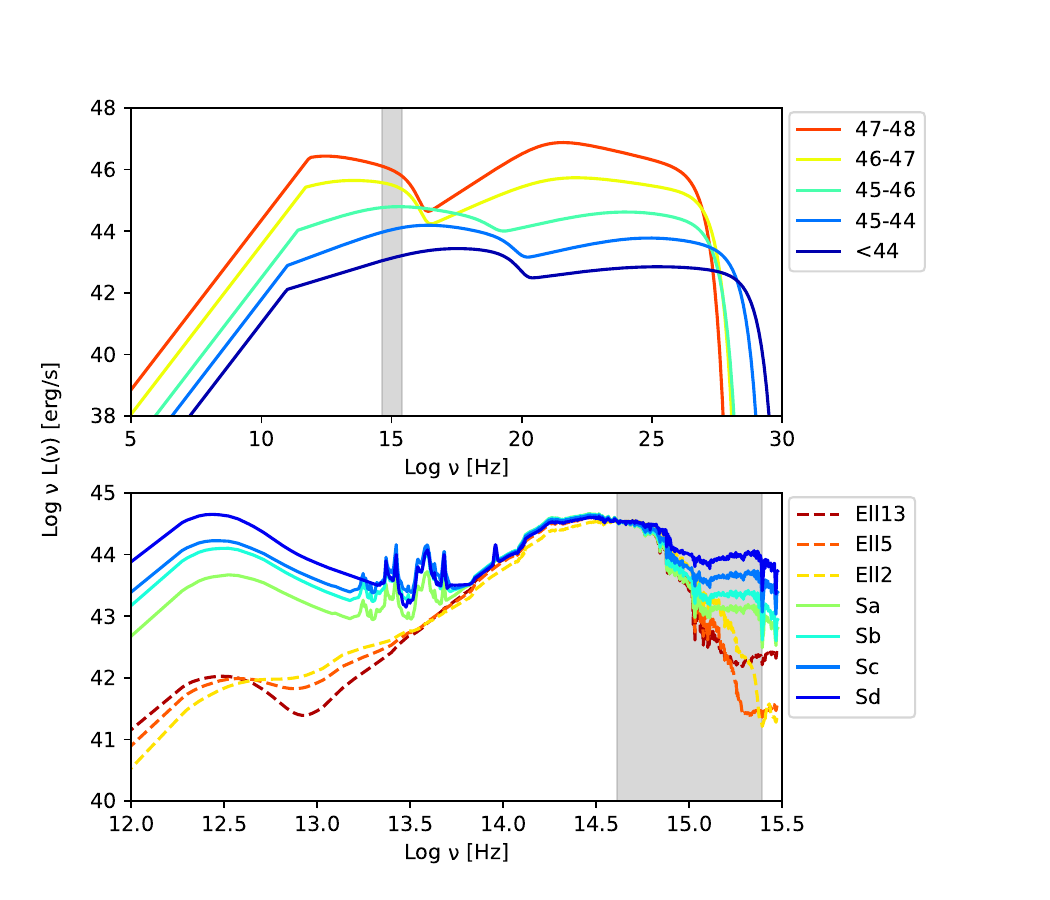}
    \caption{\textit{Top}: phenomenological SEDs for the 5 $\gamma$–ray luminosity bins derived for BL Lacs by \cite{ghisellini17}, as labelled.
    \textit{Bottom}: spiral (solid lines) and elliptical (dotted lines) galaxy templates from \cite{Polletta_2007} used to model the host galaxy emission in the simulated spectra.
    The grey shaded area in both panels represents the working range of the QSFit software, between $1215$ {\AA}  and $7300$ {\AA} (rest frame). }
    \label{fig:plot_composito_BLLacs_hosts}
\end{figure}

The simulated spectra were created in the UV to near-IR frequency range. They are a combination of phenomenological BL Lac SEDs of 5 different $\gamma-$ray luminosities from \cite{ghisellini17}, and 7 host galaxy templates from the Swire\footnote{\url{https://www.iasf-milano.inaf.it/~polletta/templates/swire_templates.html}} collection by \citet{Polletta_2007}. These 12 elements are shown in Figure \ref{fig:plot_composito_BLLacs_hosts}.
We considered a set of redshift values between 0.1 and 2, equispaced by 0.1. 

The host galaxy templates include 3 elliptical galaxies with ages of 2, 5, and 13 Gyrs, and 4 spiral galaxies of different classifications: Sa, Sb, Sc, Sd. Their absolute magnitudes vary in the range $M_R=[-21.3, -24.3]$, consistently to the distributions obtained by \cite{Sbarufatti_2005,shaw13,pita14,kasai23}. 
The jet components are derived from the phenomenological SEDs by \cite{ghisellini17}, taking into account all the BL Lac $\gamma$-ray luminosity bins reported in the original paper. 
To track the host-to-jet ratio of our synthetic spectra, we re-normalised all the SEDs in the $R$-band. 
The distribution of hosts and jets luminosities  are reported in Table \ref{table:luminosities}.

To take into account different jet behaviours or possible slightly misaligned blazars, we built a supplementary set of jet SEDs with features differing from the standard sequence. 
To implement them, we re-normalize the lowest frequency peaked BL Lac SEDs to the fainter luminosity bin (closer to misaligned intermediate BL Lacs), and the highest frequency peaked ones to the most luminous bin (closer to blue FSRQs or masquerading BL Lacs in the UV to near-IR range).  
This supplementary set includes 1400 synthetic spectra.

In order to account for observational noise in the synthetic spectra, a Gaussian noise was introduced on the flux density ($F_\lambda$).
For each point at a given wavelength $\lambda$ of the spectrum, a normally distributed perturbation was applied, with a standard deviation $\sigma_\lambda=10\%F_\lambda$. 
The noise was then extracted from Gaussian distributions centered on each flux density value, with the respective standard deviations $\sigma_\lambda$.
The uncertainties on the flux density values were defined as $\sigma_\lambda$.
This procedure was applied to the entire collection of synthetic spectra to simulate the effect of observational noise to a first approximation.
To ensure that our choice of a 10\% noise level does not introduce biases or significantly affect the overall results, we also conducted tests with noise levels of 1\%, 5\% and 15\%. While some little differences were observed, the resulting spectral behavior remained unchanged (see Section \ref{sec:confusion} for a detailed discussion of the noise effect on the spectroscopic analysis).

Note that in the case of a real spectrum, noise amplitude and flux uncertainties depend significantly on the instrument involved, its features, the observing set up and conditions. 
We chose values that are not dissimilar from widely used data sets, such as those from the Sloan Digital Sky Survey \citep[SDSS, ][]{york00}, but the results of our approach might be different (likely more effective) if high-resolution spectroscopic facilities are used.

\begin{table}[h!]
\centering
\caption{Input host magnitudes (first column) and $\gamma$-ray luminosities (third column) for the synthetic spectra, with the respective $R$-band luminosities used for defining the host-to-jet ratios (second and last columns). 
}
\label{table:luminosities}
\begin{tabular}{c c | c c}
\hline
 $M_R$ & $L_R^{\rm host}$ & $L_\gamma$ & $L_R^{\rm BL Lac}$   \\
       & erg/s            &  erg/s     &  erg/s \\
\hline
\hline
$-21.3$   & $7.50\times10^{43}$  & $<10^{44}$           & $6.34\times10^{42}$   \\
$-22.05$  & $1.50\times10^{44}$  & $10^{44}-10^{45}$    & $5.10\times10^{43}$   \\
$-22.9$   & $3.27\times10^{44}$  & $10^{45}-10^{46}$    & $4.17\times10^{44}$   \\
$-23.55$  & $5.96\times10^{44}$  & $10^{46}-10^{47}$    & $4.44\times10^{45}$   \\
$-24.3$   & $1.19\times10^{45}$  & $10^{47}-10^{48}$    & $2.10\times10^{46}$   \\
\hline
\end{tabular}
\end{table}
By combining host galaxy magnitudes and templates, and $\gamma$-ray luminosities of the blazar jet across different redshifts, we obtained a sample of 3500 synthetic spectra in the wavelength range between $1000$ and $10^7$ \AA.  

We then proceeded to implement a spectroscopic analysis approach in order to determine the conditions under which the identification of the host galaxy can be reliably obtained.

\section{Spectroscopic analysis}
\label{section:3}
QSFit\footnote{\url{https://github.com/gcalderone/QSFit.jl}} is a public software built to perform an automatic analysis of QSOs optical spectra \citep{calderone17}. 
We chose to use this software to allow the community to easily apply our analysis.
To replicate the QSO emission profile, QSFit includes a continuum component modelled as a single power-law, a host galaxy component extracted from the Swire repository (Ell5 by default), and a complete set of broad and narrow emission lines typical of quasars, including iron and Balmer pseudo-continua.
The basic QSFit \texttt{recipe} does not fit our science case, that is focussed on BL Lacs, and therefore we had to modify it. 

Since the spectrum of BL Lacs typically has very faint emission lines, and sometimes no emission lines at all, we removed all the emission lines from the standard QSFit \texttt{recipe}, i.e.\ all the known broad and narrow emission lines, the UV-optical emission due to iron, Balmer emission lines and the nuisance lines.

In our case the continuum component is expected to represent the jet BL Lac emission. The top panel of Figure \ref{fig:plot_composito_BLLacs_hosts} shows that BL Lacs with different $\gamma$-ray luminosities have different slopes and profiles in the QSFit range. For this reason, we first considered both a single  and a smoothly broken power-law. However, the wavelength range considered here is rather narrow, hence the tests on a small sample of synthetic spectra showed that the broken power-law has no significant advantage in place of a single power-law. We thus decided to stick to the default continuum profile. 

Finally, we set up two spectral analyses \texttt{recipes} characterised by two different host galaxy templates: the 5 Gyr old Elliptical and the b-type Spiral galaxy (Ell5 and Sb in the Swire repository, respectively).
Every \texttt{recipe} represents a different spectroscopic analysis, through which one can obtain the estimate of the power-law continuum and host galaxy template that better describe an input spectrum\footnote{
The recipes to replicate our analysis are available at \url{https://github.com/gaiadelucchi/BL-Lac-host-recipe}.
}.

For each of the 3500 spectra, we performed two analyses, one for each \texttt{recipe}. 
Every QSFit run with the modified \texttt{recipe} returns best fit values in a data structure called \texttt{res}, including the fit statistics, the continuum and the host galaxy components, 
(\texttt{res.fitstats}, \texttt{res.bestfit.params[:QSOcont]} and \texttt{res.bestfit.params[:Galaxy]}), that we detail in the following.
We underline that, even if the QSFit parameter related to the single power-law is referred to as \texttt{QSOcont}, in our case it refers to the non-thermal jet emission component. The naming is inherited from the original QSFit purpose. 
In detail, the \texttt{res} parameters are:
\begin{itemize} 
    \item[-] \texttt{res.fitstats.fitstat}: \\the reduced chi-square ($\chi_{red}^2$) obtained from the QSFit minimization procedure;
    \item[-]\texttt{res.bestfit.params[:QSOcont].x0}: \\a wavelength fixed value located in the middle of the available spectral wavelength range, to define the power-law;
    \item[-]\texttt{res.bestfit.params[:QSOcont].norm.val} \\normalisation of the continuum component at \texttt{x0}, in units of $10^{42}$ erg s$^{-1}$; 
    \item[-]\texttt{res.bestfit.params[:QSOcont].norm.unc}: \\uncertainty on the power-law normalisation, in the same units; 
    \item[-]\texttt{res.bestfit.params[:QSOcont].alpha.val}: \\the slope of the power-law, constrained in the range [-3,1];
    \item[-]\texttt{res.bestfit.params[:QSOcont].alpha.unc}: \\uncertainty on the continuum slope;
    \item[-]\texttt{res.bestfit.params[:Galaxy].norm.val}: \\value of the host galaxy normalization at 5500 \AA, in units of $10^{42}$ erg s$^{-1}$; 
    \item[-]\texttt{res.bestfit.params[:Galaxy].norm.unc}: \\uncertainty on the host galaxy normalisation, in the same units.
\end{itemize}

\subsection{Analysis results}

QSFit allows us to decompose synthetic spectra obtaining numerous parameters useful to determine the goodness of the fit. For both  \texttt{recipes} we focus on three parameters for our study: the fit statistic ($\chi_{red}^2$), the power-law normalisation and host galaxy normalisation,  with the relative uncertainties.
For a more uniform definition, we redefine the power-law normalisation by calculating it at 5500\AA\ instead of \texttt{x0} as defined by QSFit.

The input parameters used to build the synthetic spectra, the galaxy templates used to perform the QSFit analysis and the fit parameters mentioned above are the values crucial to define a method to 
directly identify the host galaxy class in a BL Lac spectrum.

\begin{figure}[h!]
\includegraphics[width=9cm]{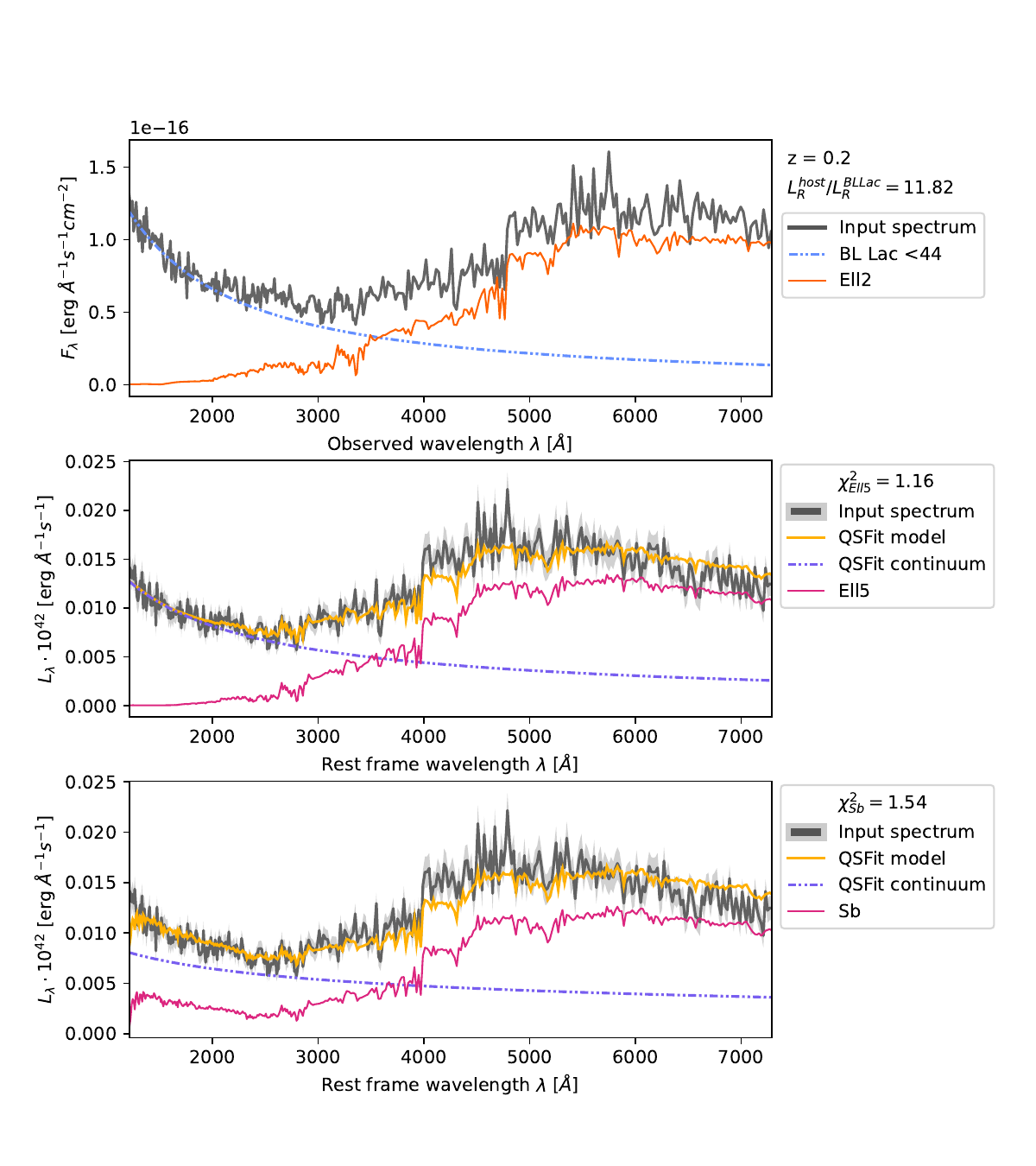}
\caption{\textit{Top}: synthetic spectrum with PL and host galaxy components comparable, with host-to-jet ratio of the order of unity as labelled. 
\textit{Middle, Bottom}: QSFit spectral analysis performed on the spectrum of the top panel, respectively with Ell5 and Sb.
In this example, QSFit recognises the correct family to which the host galaxy belongs: the $\chi^2_{red}$ obtained with the elliptical template Ell5 is better than the one obtained with spiral Sb.}
\label{fig:Ell2_comparable_OK}
\end{figure}

\begin{figure}[h!]
\includegraphics[width=9cm]{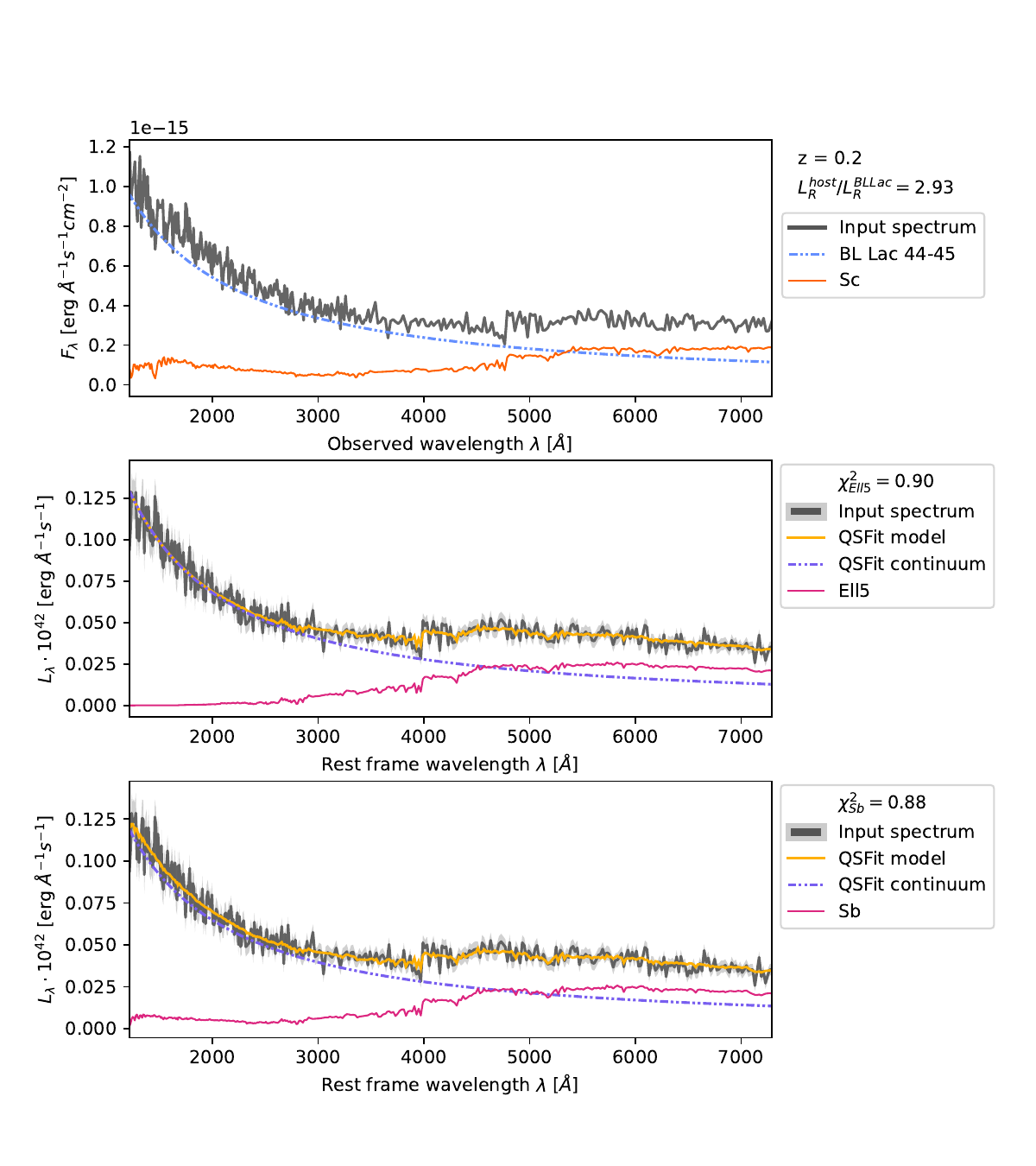}
\caption{\textit{Top}: synthetic spectrum with PL and host galaxy components comparable, with host-to-jet ratio of the order of unity as labelled. 
\textit{Middle, Bottom}: QSFit spectral analysis performed on the spectrum of the top panel, respectively with Ell5 and Sb.
Here the noise in the bluer region of the spectrum doesn't allow QSFit to distinguish the correct type of the host galaxy, as we can see from the $\chi^2_{red}$ of analyses.}
\label{fig:Sc_comparable_NO}
\end{figure}

\begin{figure}[h!]
\includegraphics[width=9cm]{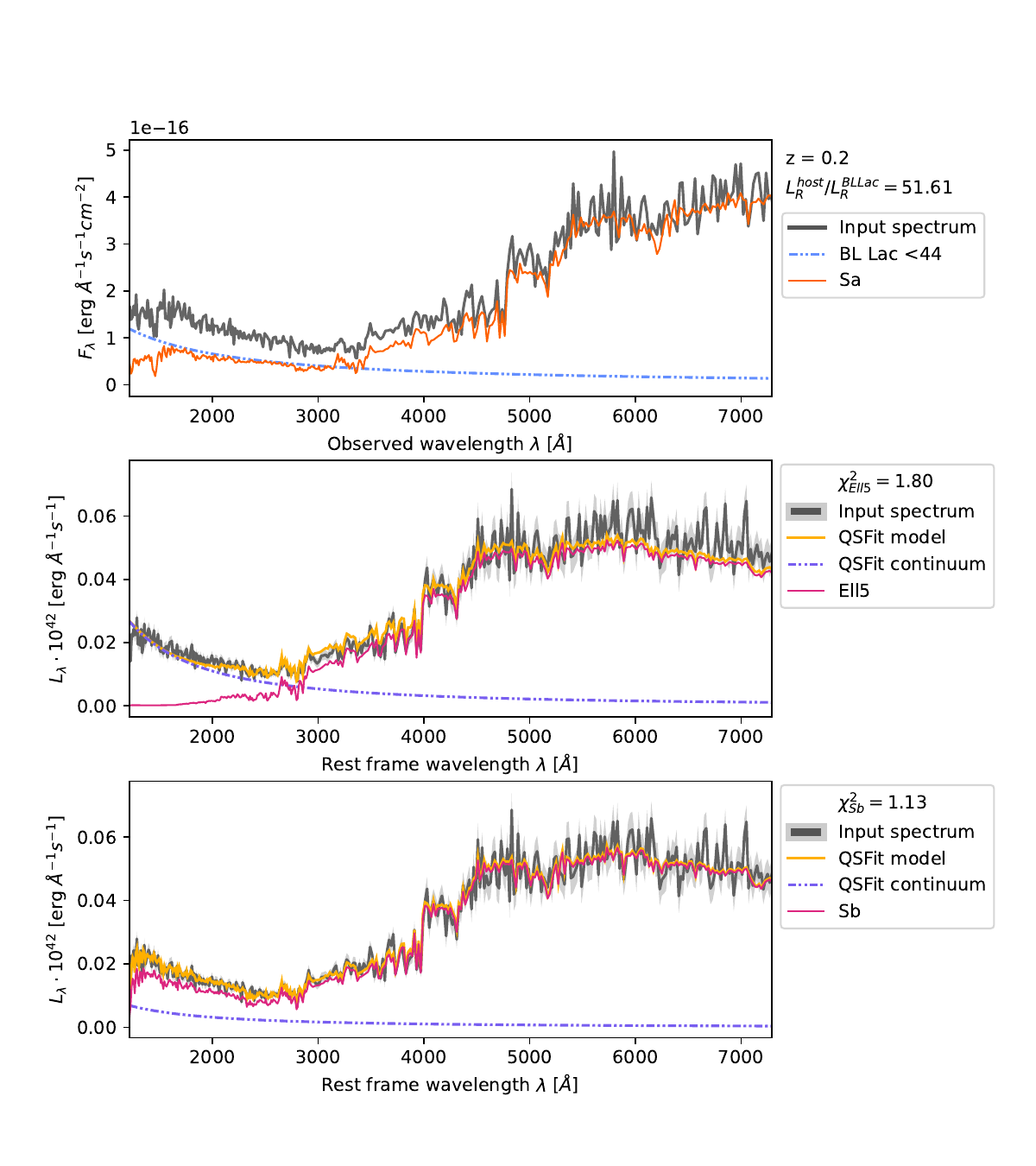}
\caption{\textit{Top}: synthetic spectrum with dominant host galaxy component, with a large host-to-jet ratio, as labelled.
\textit{Middle, Bottom}: QSFit spectral analysis performed on the spectrum of the top panel, respectively with Ell5 and Sb.}
\label{fig:Sa_dominant}
\end{figure}

Considering the R-band luminosity of the hosts and the BL Lac jets, we expected to have three cases:   

\begin{itemize}
    \item[-]$L_R^{host} \ll L_R^{BLLac} $: the host galaxy luminosity is negligible compared to the jet luminosity, so we are not supposed to distinguish the host galaxy in the spectrum, being completely diluted by the BL Lac emission. QSFit, in this case, is expected to provide a continuum-dominated fit and a host normalisation which is compatible with zero,  regardless of its type.
    \item[-]$L_R^{host} \sim L_R^{BLLac}$: the luminosities of the two components are comparable, the host galaxy is expected to be visible and the distinction between ellipticals and spirals can become appreciable depending on the host-to-jet ratio. This case is shown in Figures \ref{fig:Ell2_comparable_OK} and \ref{fig:Sc_comparable_NO}. 
    The top panel shows a synthetic spectrum with its components, the middle and bottom ones show the QSFit analysis performed with the Ell5 and Sb \texttt{recipes}, respectively. These last two plots show the original synthetic spectrum with its uncertainties in shades of grey, and the QSFit model with its individual components in coloured, thinner lines. 
    Figure \ref{fig:Ell2_comparable_OK} shows clearly that the analysis performed with the Sb galaxy template is worse than the Ell5-based one, 
    as also suggested by the reduced chi-squared statistics.
    Specifically, the bluer part of the spectra shows the most significant differences: since spiral galaxies are bluer than ellipticals due to star formation, the region at shorter wavelengths is crucial to determine the classification of the hosts. When QSFit tries to overlap a spiral galaxy on an elliptical one, due to this region, the resulting goodness of the fit is lower than fitting with a different elliptical template (and vice-versa).
    It is important to disclaim that the simulated Gaussian noise is proportional to the flux density, thus affecting more significantly the brightest parts of the spectrum. 
    Input templates with brighter blue-UV regions, e.g.\ Sc spirals as in Figure  \ref{fig:Sc_comparable_NO}, might thus be penalised in their fit statistics, preventing QSFit from correctly recognising the host. 
    As the host-to-jet ratio increases, QSFit gets better at host galaxy identification, despite the definition of the noise.
    \item[-]$L_R^{host} \gg L_R^{BLLac}$: the host galaxy dominates the jet emission and is expected to be more clearly recognizable. In this case the host-to-jet ratio is larger than 1, the features of the galaxy template become dominant with respect to the blazar power-law. For this reason, the difference between spirals and ellipticals at shorter wavelengths grows as the ratio increases. 
    This is the case of Figure \ref{fig:Sa_dominant}.
    The top panel shows the input spectrum with Sa as a host. 
   The middle panel clearly shows that the model generated by the QSFit analysis using the Ell5 template lacks features between $1215${\AA} and $2000${\AA} and the reconstructed luminosity density in the longer wavelength part is lower than the spectrum. In contrast, the last panel shows the fit performed with the Sb template, which shares some features with the Sa spiral at the shorter wavelengths. These common features allow to constrain the bluer region of the spectrum and thus also better reconstructs the redder region.
    This is confirmed also by the resulting statistics obtained from the spectral analysis: with the elliptical galaxy template QSFit gives $\chi^2_{Ell5} = 1.80$, whereas analysing the same spectrum with spiral it gives $\chi^2_{Sb} = 1.13$, which is closer to 1.
\end{itemize}

As the host-to-jet ratio increases and the galaxy becomes more visible, the characteristic features of elliptical or spiral galaxies are expected to become easier to distinguish.
In order to understand which galaxy template between elliptical and spiral best fitted the spectrum, it is possible to use the statistic reported by the QSFit results. 
A significant difference in the $\chi_{red}^2$ value is expected if an input spectrum is fitted with a  \texttt{recipe} including a host galaxy template of the same galaxy type, or if instead the galaxy type of the input and the \texttt{recipe} are different. We applied systematically this approach to all the synthetic spectra in order to define which was the combination of power-law and host galaxy template that best fitted the synthetic spectrum. 
We thus introduce in the following section a new approach to automatically distinguish the host galaxy nature of an observed BL Lac, based on the difference in statistics obtained by just two spectroscopic analyses. 

\section{How to identify the host galaxy}
\label{section:4}

All simulated spectra were analysed twice, once with the \texttt{recipe} including Ell5 as a host template, once with the Sb, as described in Section \ref{section:3}.
In order to understand which galaxy template was best fitted, we took into account the statistics provided by the two QSFit analyses. 

As mentioned above, the $\chi_{red}^2$ obtained by the analysis performed with the galaxy template most similar to the input galaxy is expected to be better. 
In other words, spectra with elliptical (spiral) galaxies in input are expected to yield a $\chi_{red}^2$ closer to 1 when analysed with the Ell5 (Sb) template. 

Following this logic, we should be able to infer the galaxy type in the analysed spectrum using the ratio of the two statistics, defined as: 
\begin{equation}
    R=\frac{\chi^2_{Ell5}}{\chi^2_{Sb}}.
\end{equation}
We identify 3 different scenarios:
\begin{itemize}
    \item[-]$R<1$: QSFit provides a better fit with elliptical template Ell5, this case corresponds to $\chi^2_{Ell5}<\chi^2_{Sb}$;
    \item[-]$R=1$: QSFit provides comparable statistics for both host galaxy templates, in this case $\chi^2_{Ell5}=\chi^2_{Sb}$ so we are not able to distinguish which is the host galaxy;
    \item[-]$R>1$: QSFit provides a better fit with the spiral template Sb, it corresponds to $\chi^2_{Ell5}>\chi^2_{Sb}$.
\end{itemize}

\begin{figure*}[h!]
    \includegraphics[width=0.95\textwidth]{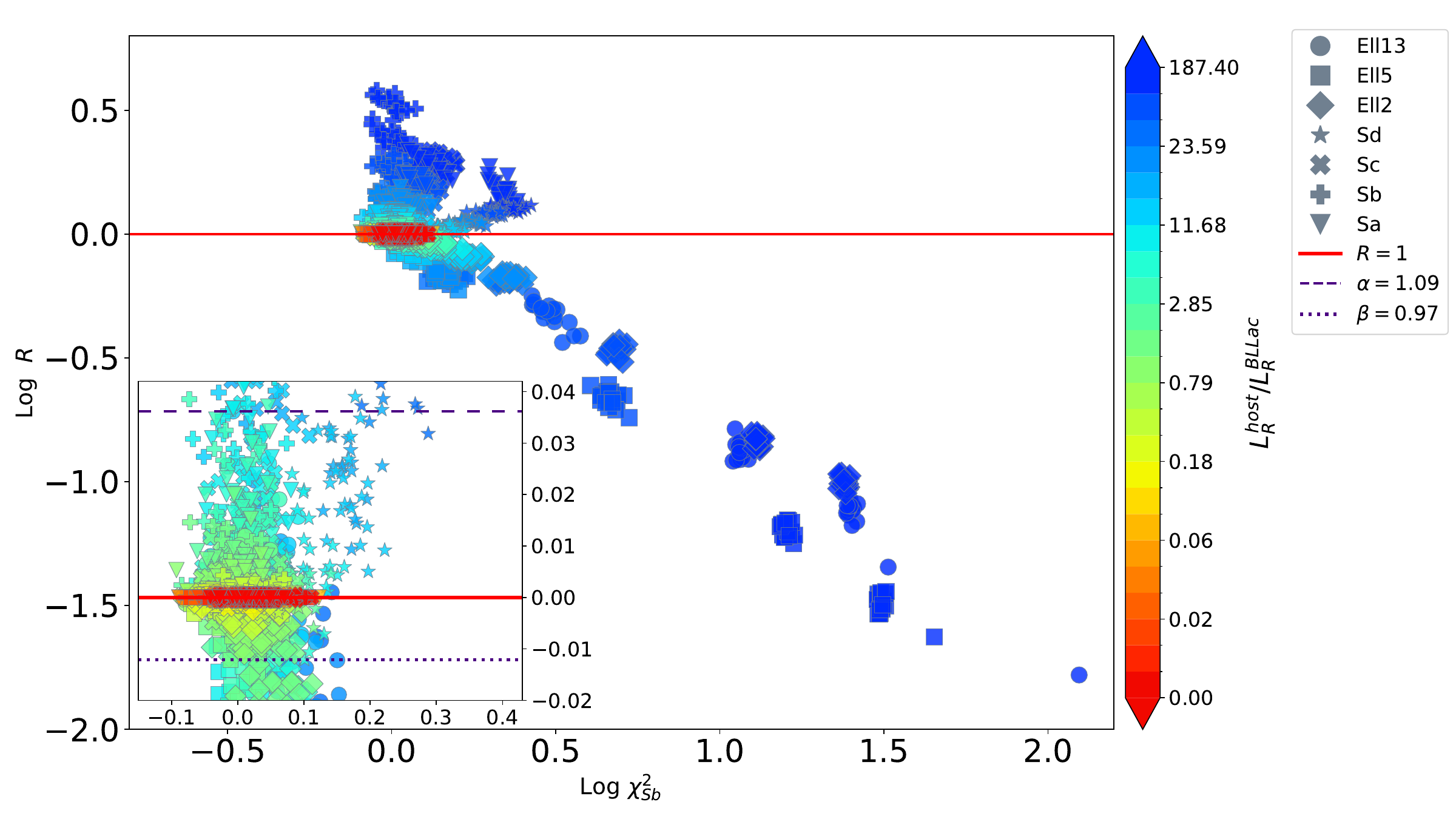}
    \includegraphics[width=0.75\textwidth]{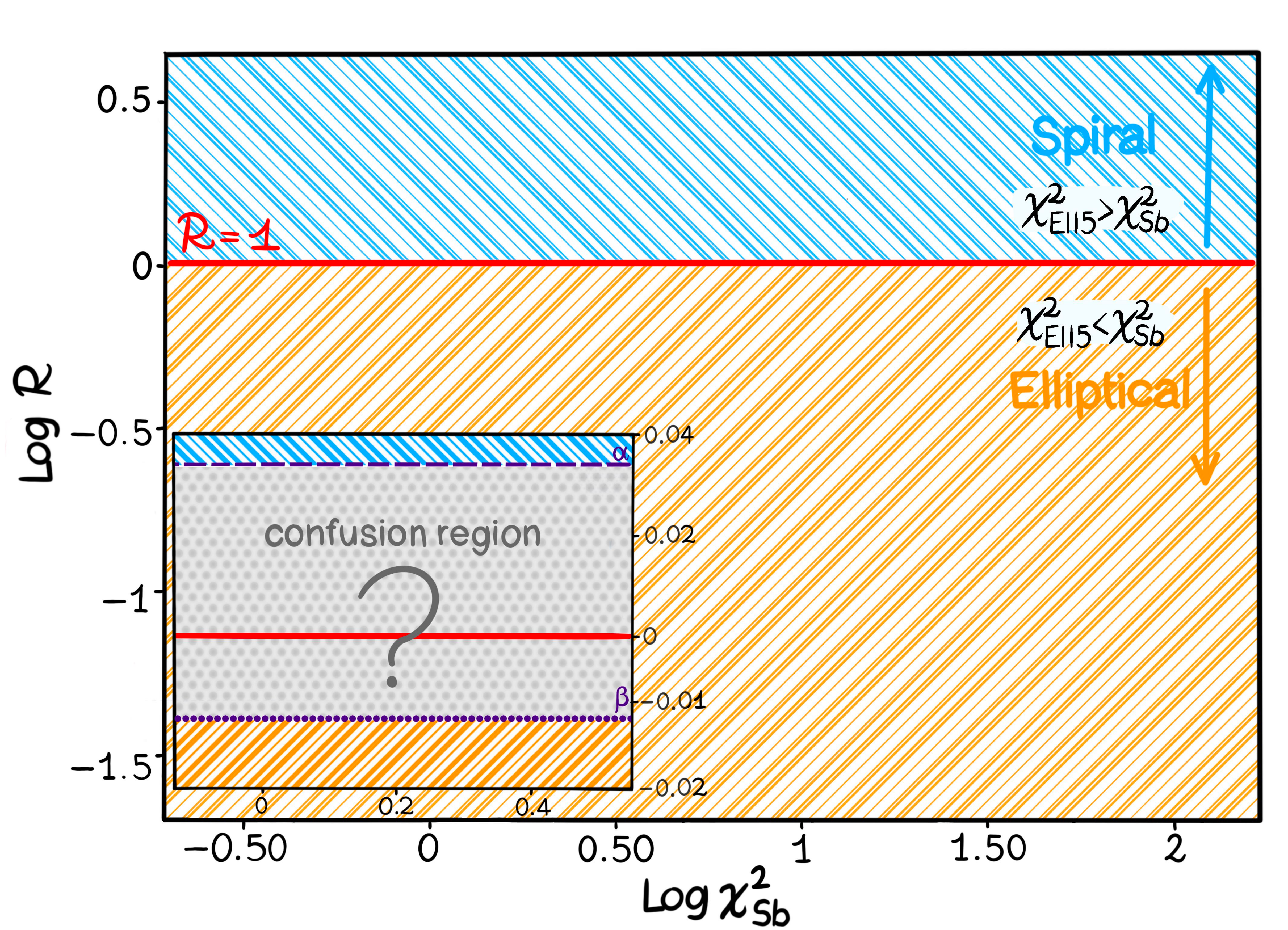}
    \caption{\textit{Top}: the plot shows the ratio of the statistics $R$ obtained from the two analyses performed by QSFit with the two \texttt{recipes}. The different colors indicate the R-band host-to-jet ratio ($L_R^{host}/L_R^{BLLac}$) as shown with the colourbar, the markers shapes refer to the input galaxy templates of the spectra. The linear fit in red was obtained by fitting the points of the spectra that obtained null host galaxy normalisation from both QSFit  \texttt{recipes}, i.e. those with dominant jet and thus same statistics for both galaxy templates of the QSFit analyses.
    The inset of the plot shows the zoom of the region around $R=1$, the purple dashed lines indicates the upper ($\alpha$) and the lower ($\beta$) limits of the region of uncertainty.
    \textit{Bottom}: the sketch shows how we can use the previous plot to discriminate between spiral and elliptical host galaxies in our BL Lac synthetic sample. The confusion region is the area where we cannot distinguish whether the host galaxy is elliptical or spiral.} 
    \label{fig:plot_R}
\end{figure*}

In the following, we will use Figure \ref{fig:plot_R} as a reference to set up our method to discriminate between spiral and elliptical galaxy hosts in our BL Lac synthetic sample.
The top plot in this Figure shows $R$ as a function of the $\chi^2_{Sb}$ obtained by fitting the spectra with the \texttt{recipe} defined with the Sb galaxy template. 

Considering the host-to-jet luminosity ratio $L_R^{host}/L_R^{BLLac}$ shown by the colourbar on the right of the plot, we can observe that the points are distributed in an orderly way according to the luminosity ratio. In fact, the points belonging to the synthetic spectra with negligible host luminosity compared to the blazar luminosity ($L_R^{host} \ll L_R^{BLLac}$), and therefore with a luminosity ratio close to zero, lie very close to the red line indicating $R=1$, and QSFit provides the same values regardless of the host galaxy template being considered.
In this case, distinguishing the host galaxy is not possible as it is completely diluted by the intense blazar jet emission. 
Moving away from the red line, and then towards higher luminosity ratios, the host galaxy becomes visible in the spectra ($L_R^{host} \sim L_R^{BLLac}$), and for some of them it is now possible to clearly distinguish the type of galaxy. 
Finally, for points associated with spectra with host galaxy luminosity significantly brighter than the jet luminosity ($L_R^{host} \gg L_R^{BLLac}$), the points move further away from the line and the host galaxy type of the synthetic spectrum becomes more distinguishable. 

It is thus possible to identify a quick criterion to distinguish which type of host galaxy was better fitted by QSFit by using the red line as a discriminant: above the line the host is more likely a spiral, below it is more likely an elliptical.
In its proximity instead we cannot clearly distinguish the host galaxy family to which it belongs (insets and sketch in Figure \ref{fig:plot_R}). 

\subsection{The confusion region}
\label{sec:confusion}

When the statistics obtained with the two spectroscopic analysis are comparable (i.e. $R\simeq1$,  within 1\%), contamination from the "wrong" host galaxy template can occur on both halves of the plot.  This happens in particular for small host-to-jet ratios.

The following discussion is based on the case with 10\% noise, and the core set of synthetic spectra. Results for the other explored noise levels and the supplementary ``out-of-sequence" SEDs will be detailed later.

In order to precisely assess the contamination, we introduced two limits on the $R$ value:
\begin{itemize}
    \item[-]an upper limit $\alpha=1.09$, which corresponds to the maximum value $R$ assumed by an elliptical contaminant in the spiral region ($R>1$);
    \item[-]a lower limit $\beta=0.97$, the minimum value $R$ assumed by a spiral contaminant in the region identifying elliptical hosts ($R<1$).
\end{itemize}

These limits define the area $\beta<R<\alpha$ within which we cannot infer the type of host galaxy just by following our two-fit-approach. We called this area \textit{confusion region}. 

Within this region ($\beta < R < \alpha$), host galaxy misclassification can occur in both directions:
\begin{itemize}
    \item[-]in the range $1<R<\alpha$, elliptical host galaxies are incorrectly classified as spirals;
    \item[-] in the range $\beta<R<1$, spirals hosts are missclassified as ellipticals.
\end{itemize}

The 70\% of all sources fall within the confusion region ($\beta < R < \alpha$), highlighting the need to interpret classifications in this range with caution. Among them, 51\% have $R$ values consistent with 1 at 1\% level, further confirming the difficulty in distinguishing the host galaxy type for these cases.
However, these two types of contamination are not equally significant, as can be seen from the asymmetry of the confusion region around $R=1$ in Figure \ref{fig:plot_R}. 
Several spectra with elliptical hosts fall erroneously in the spiral region, showing $R$ values consistent with those derived in the case of spectra simulated with spiral templates. The elliptical contaminants of the spiral region are mainly created with Ell13 galaxy template. The Ell13 galaxy template in the Swire collection shows an excess flux in the blue-UV range compared to Ell2 and Ell5. As said above, ellipticals and spirals share features in the visible range at wavelengths corresponding to the red and near-IR, so they might be confused by studying only the reddest part of the spectrum: this excess flux in the bluest part could justify the inability to distinguish Ell13, especially when added to the jet power-law. 
On the other hand, the number of misidentified spirals falling in the elliptical region is roughly negligible because their $\chi^2_{red}$ ratio remains consistent with $R=1$ at less than a 1\% level. A more detailed view of the asymmetric $R$ distribution between elliptical and spiral templates is shown also in the left panel of Figure \ref{fig:R_withshiftedSEDs} in the Appendix.

We chose a medium age elliptical and an intermediate spiral template to perform our double fitting and set up the classification method. Specific host galaxies may have features that an automatic analysis as in QSFit may confuse: the excess blue-UV brightness in the Ell13 template can be easily confused with typical spiral signatures by the spectroscopic analyser, and some features that differentiate spiral galaxies can sometimes be better fitted using an elliptical galaxy template.
This is the main reason for a systematic approach in studying blazar host galaxies, and for a conservative definition of the confusion region.


It should be noted that without considering the outlier, the amplitude corresponding to $\alpha$ is approximately 4 times that of $\beta$, and the contamination in the elliptical region (i.e. $R<1$) is thus almost negligible. Most of spiral-hosted BL Lacs that populate the lower end of the confusion region are strongly dominated by jet emission (for which QSFit assigns a null host galaxy normalisation). The same cannot be said for contamination in the spiral region.

When considering the supplementary set of synthetic spectra built with jet SEDs out of the \cite{ghisellini17} sequence, the results do not vary significantly.
In fact,  the upper limit $\alpha$ and the lower limit $\beta$ do not vary adding this set, and 64.2\% of all sources fall within the confusion region\footnote{Note that we consider only two extreme SEDs to test the out-of-sequence cases. With a more complete sampling the fraction of sources in the confusion region might slightly change.}. 
The complete version of Figure \ref{fig:plot_R} built on both the core sample and the supplementary out-of-sequence spectra is included in the Appendix. 


We also tested our approach using different fixed noise amplitudes in the synthetic spectra — specifically 1\%, 5\% and 15\% —  in addition to the original tests with 10\% noise level.   These tests confirm that the upper and lower boundaries of the confusion region fluctuate only slightly, as shown in Table \ref{tab:Confusion_table}. 
\begin{table}[]
\caption{Limits of the confusion region and fraction of sources falling in it, for spectra obtained with different noise levels.}
    \centering
    \begin{tabular}{c|cccc}
    \hline 
 noise level& 1\%&5\% &10\% &15\%\\
    \hline 
 $\alpha$&  1.53&1.14& 1.09&1.08\\    
         $\beta$&  0.91&0.96& 0.97&0.89\\
         \# $\beta\le R\le\alpha$&  61.2\%&66.0\%& 70.4\%&80.1\%\end{tabular}
 
    \label{tab:Confusion_table}
\end{table}

Interestingly, we observe that increasing the noise amplitude leads to a reduction in the extent of the confusion region. This is intuitively expected, since the spectra become less distinguishable and the points concentrate around $R = 1$. A closer look at the bounding parameters of the confusion region reveals that $\beta$ value remains relatively stable across different noise levels, while $\alpha$ shows a more significant variation. This different behaviour may suggest a bias in accurately distinguishing between certain morphological types, in particular against spiral galaxies.

We finally tested a different spectral resolution for our input spectra, by reducing at 1/3 the original one inherited by the Swire host galaxy templates ($R_s\sim120)$.
As we expected, the bounding limits get worse ($\alpha=1.21$, $\beta=0.88$).

A critical point is the wavelength coverage: the bluest portion of the spectra has an important role in constraining the host galaxy contribution, and there is limited availability of UV spectrographs that allow for such extended ranges. We thus tested our analysis on a limited wavelength range: limiting the analysis to a 3500-7300\AA\ range impacts on the ability to discriminate between host galaxy types. If possible, this analysis should thus be performed on spectra including a good UV coverage. 

\subsection{Luminosity limits}
\label{sec:luminosity}

An important result of our analysis is to understand for which BL Lac classes a host galaxy can realistically be studied with QSFit. 
It is immediately clear, in fact, that the synthetic spectra built with the brightest BL Lac profiles defined in \cite{ghisellini17} have too luminous relativistic jets to allow for a reliable identification of other dimmer components in the same wavelength range.
All spectra with $R=1$ have $L_R^{host} < 0.074\, L_R^{BLLac}$. 
This translates in a limit luminosity $L_R^{BLLac}\simeq4.4\times10^{45}$erg/s for a host galaxy magnitude of $M_R=-22.9$ (the peak of the distribution), above which the host galaxy is too faint to be statistically relevant in the QSFit spectroscopic analysis\footnote{
This limiting value varies according to the input host galaxy magnitude, from $2.1\times10^{46}$ to $1.3\times10^{45}$ erg/s for the input limiting magnitudes $M_R=-24.3$ and $-21.3$, respectively.
}.
In the classification by \cite{ghisellini17}, BL Lacs with $\gamma$-ray luminosities in the bins ${\rm Log}L_\gamma=46-47$ and $47-48$ fall in this group. They are the $21\%$ of the sample from which the $\gamma$-ray binning was defined. 
We expect that our approach could be successfully applied to the remaining $79\%$ of that sample. 
The same $\gamma$-ray luminosity limits are obtained when the analysis is performed on spectra with 5\% and 15\% noise amplitudes.

With an analogous approach, we derived the maximum BL Lac $R$-band luminosities below which it is consistently possible to identify elliptical ($R<\beta$) and spiral ($R>\alpha$) galaxies.
Specifically, we can correctly classify a spiral host when $L_R^{host} > 2.85\, L_R^{BLLac}$, which corresponds to a BL Lac limit luminosity $L_R^{BLLac}\simeq1.2\times10^{44}$erg/s for the mean magnitude value of the host galaxies.
We can instead identify an elliptical host when $L_R^{host} > 0.79\, L_R^{BLLac}$, which translates in a limit luminosity $L_R^{BLLac}\simeq4.2\times10^{44}$erg/s.
Ellipticals can be more easily classified than spiral host galaxies in case of brighter BL Lac jets.

While these limit luminosities are not different when derived with a 5\% and 10\% noise level applied to the synthetic spectra, a larger noise (15\%) implies slightly larger host-jet ratios to correctly distinguish different types of host galaxies, as intuitively expected.
A larger noise level and uncertainties reduce the $\chi^2_{red}$ distribution, slightly narrowing the confusion region as well, and allowing only the brightest hosts to stand out: spirals are identified when $L_R^{host} > 6.4\, L_R^{BLLac}$ and ellipticals for $L_R^{host} > 1.4\, L_R^{BLLac}$, for a 15\% noise amplitude.  

Despite a small uncertainty region, we have introduced a method to easily distinguish the host galaxy type of BL Lacs, using only two spectroscopic analyses for each synthetic spectrum.

\section{Test on SDSS BL Lac spectra}

In order to test and evaluate the reliability of our spectroscopic method for identifying and classifying the host galaxies of BL Lac objects, we employed a sample of optical spectra from the SDSS.
The sample was selected from the fourth catalog of AGN \citep[4LAC-DR3; ][]{ajello20} detected by the Large Area Telescope (LAT) onboard the {\it Fermi} satellite \citep{atwood09}, choosing the sources with $\gamma$-ray luminosities in the 0.1–100 GeV energy range below the threshold derived from the analysis on synthetic spectra ($L_{\gamma} = 10^{46} \rm erg/s$). From the catalog we generated by applying this $L_\gamma$ cut, we obtained a sample of 240 optical spectra by performing a cross-correlation with the SDSS-DR17 source list.
The galactic extinctions $A_{\rm V}$ for each source were retrieved from the NASA/IPAC Extragalactic Database (NED).
Our spectroscopic method was applied to all sources in the final data sample, each analysed twice with QSFit, following the procedure described in Section \ref{section:4}.

\begin{figure}[h!]

\includegraphics[width=8cm]{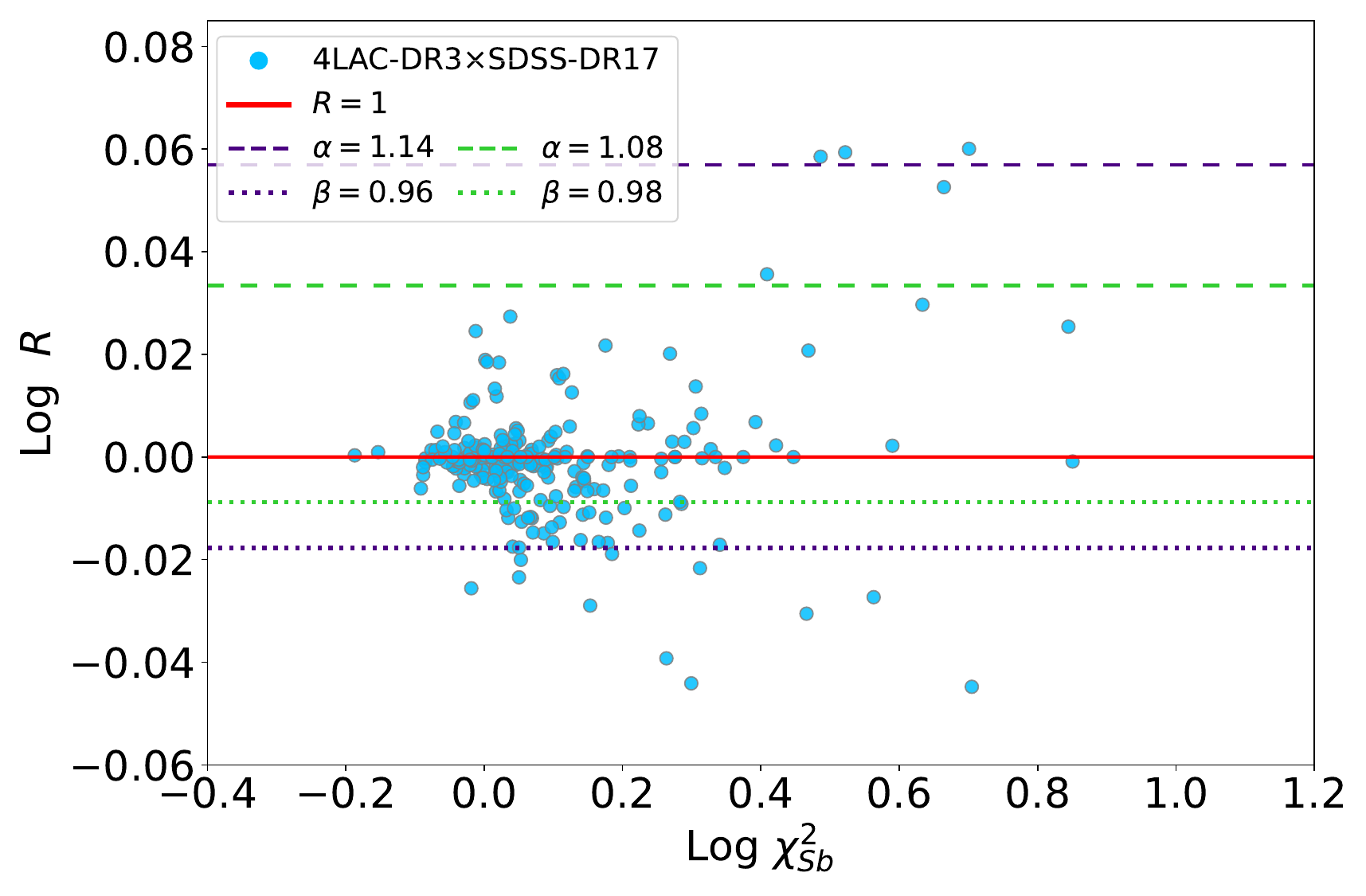}
\caption{Results of the analysis performed on a sample of BL Lacs with $L_{\gamma} \leq 10^{46} \rm erg/s$. 
         The plot is structured as Figure \ref{fig:plot_R}. 
         Only 14 sources fall outside the confusion region defined for 5\% noise spectra (limits in purple), and all in the elliptical section. 
         The green lines show the limits of 5\% tails. 
         }
\label{fig:SDSS_data}
\end{figure}
Since the median signal-to-noise ratio of the SDSS spectra in this sample is 20.7, we chose to compare the QSFit results with those obtained from the analysis of 5\%-noise synthetic spectra. 

According to those limits, 14 out of 240 BL Lacs in our sample fall outside the confusion region. 
Only three of them fall in the spiral half of the plot. 
The other 11 fall in the elliptical half.

We also checked the sources that form the 5\% tails of the $R$ distribution, to investigate the nature of the BL Lac hosts falling closest to the confusion region limits. 
We checked their SDSS photometry and looked for possible morphological classification or relevant information in the literature.
At the lowest $R$ values, sources appear to be hosted in elliptical galaxies, as expected from our method. 
Notably instead, the 5 sources with higher $R$ values, representing the 2.5\% tail of the population on the spiral side of Figure \ref{fig:SDSS_data}, have a variety of host features:
three of them appear to be hosted in irregular or interacting galaxies, or close galaxy clusters, while the other two have clearly elliptical hosts, with no relevant interaction. 
In particular, 
SDSS J083724.50+145822.1 is hosted in an interacting galaxy, part of a rich cluster \citep{hao10},
SDSS J130248.70+475510.6 is in a rich galaxy group, likely interacting on a large scale \citep{tempel17}, and
SDSS J144248.10+120039.9 is interacting with an elliptical galaxy at an 8kpc projected distance \citep{sbarufatti06}.

This interestingly reinforces the validity of our method: the sources with a bluer stellar population tend to fall toward the spiral region of our classification plot. 
We expect that the same analysis on spectra with a wider wavelength coverage would allow to better constrain the host nature, following our approach. 
We plan to expand these observational tests (including dedicated follow-up observations and more extended BL Lac samples) in an upcoming publication.

\section{Conclusions}

In order to shed light on the nature of galaxies hosting BL Lac objects, we developed a simple method based on optical-NIR spectroscopic analysis to systematically distinguish whether the host galaxy is elliptical or spiral.
To implement and test the algorithm we first created a sample of synthetic spectra by combining different host galaxy templates and analytic BL Lac SEDs, with different luminosity combinations and located at redshifts in the range $z=0.01-2$.  
The final sample is composed by 3500 synthetic spectra.

We then modified the QSFit software to fit all spectra with simple assumptions, consistent with their BL Lac nature, i.e. blazars without emission lines. 
We defined two QSFit \texttt{recipes} characterised by two different host galaxy templates (Ell5 and Sb from the Swire template collection) 
added to the single power-law continuum built in the software standard set up. 
Each spectrum was then analysed with both \texttt{recipes}, to test QSFit capabilities and limitations in this context, and implement a classification algorithm. 
The main parameters used for classification purposes are the statistics of each fit
$\chi^2_{Ell5}$, $\chi^2_{Sb}$, and their combination $R=\chi^2_{Ell5}/\chi^2_{Sb}$.
The reference value used to interpret our results and highlight the capabilities of our method is the host-to-BL Lac input R-band luminosity ratio $L_R^{\rm host}/L_R^{\rm BL Lac}$.

We can conclude that, even assuming somewhat large noises, it is possible to discriminate the nature of BL Lac host galaxies when their relativistic jets emit at $\gamma$-ray luminosities below $\sim10^{46}$erg/s, and a spectrum in the wavelength range 1215-7300\AA\ is available\footnote{
Without a good coverage of the bluer part of this range, the possibility to distinguish between the two host galaxy types is reduced, allowing for easier biases. 
}. 
About 80\% of BL Lacs with redshift in the 3LAC sample have luminosities that allow to apply this method.

After having carried out just two spectroscopic analyses with the two adapted QSFit \texttt{recipes} on the available BL Lac spectrum, the $R$ value can give an immediate indication of the host galaxy nature:
\begin{itemize}
\item[-] if $R<\beta=0.97$ the host galaxy is most likely an elliptical
\item[-] if $R>\alpha=1.09$ the BL Lac is most likely hosted in a spiral galaxy
\item[-] if $\beta\le R\le\alpha$ the host nature is uncertain.
\end{itemize}

In general, the host galaxy classification method based on spectroscopic analysis performed with QSFit is more reliable the stronger the host galaxy component. This is also intuitively expected. 
However, we noticed that the amplitude of upper limit $\alpha$ of the confusion region is 4 times larger than the one of lower limit $\beta$.
This leads to an asymmetry in the uncertainty region that suggests caution. 
The contamination of misclassified spiral hosts in the elliptical region is almost negligible, and only amounts to the brightest relativistic jets, i.e.\ a measured host-to-jet ratio close to zero. On the other hand, there are a number of misclassified elliptical galaxies that contaminate the region of spirals that do not fall in the brightest BL Lac classes and are more sparsely distributed.
This asymmetry implies that faint spirals with respect to their jets are more easily not automatically identified, since they fall in a wider confusion region. 

Outside the narrow confusion region, our approach allows to consistently discriminate between elliptical and spiral hosts with different stellar population ages and morphologies. 

We performed a first test on BL Lacs from {\it Fermi} 4LAC-DR3 and SDSS-DR17 catalogs, that confirmed the validity of our method.

We thus expect to successfully apply it to existing BL Lac spectra and the better quality data that will be produced by new generation spectroscopic facilities, such as JWST and {\it Euclid}. 
In case of \textit{Euclid}, the Near-Infrared Spectro-Photometer (NISP) could be used also to obtain spatial information from 2D-spectroscopy. In fact, although a spectral resolution of $R_s\sim450$, for a number of sources would be possible to disentangle the host galaxy from the blazar jet contribution. The spatial resolution of NISP will be enough to separate the spectra of these two components: at the centre the jet pointing toward us, and at the edges two weaker spectra of the host galaxy.
In future, a method which combines 1D and 2D spectroscopy will be studied.






\appendix
\section{Results including out-of-sequence synthetic spectra}

The phenomenological SEDs describing the BL Lac jet emission were implemented following \cite{ghisellini17}, that discuss the behaviour of blazar SEDs in the context of the {\it Fermi} blazar sequence. 
To test our approach in case of outliers, i.e.\ blue FSRQs, masquerading BL Lacs, or misaligned intermediate frequency peaked BL Lacs, we implemented a supplementary set of jet SEDs. 
We built on the existing phenomenological SEDs and normalized (i) the reddest BL Lac SED to the fainter luminosity bin, and (ii) the bluest BL Lac SED to the brightest luminosity bin. 
This supplementary set includes 1400 new synthetic spectra, with the same redshift and host magnitude values, and same host galaxy types as the core sample. We assigned to this set of spectra a 10\% level noise. 

\begin{figure}[h!]
    \centering

    \includegraphics[width=0.46\columnwidth]{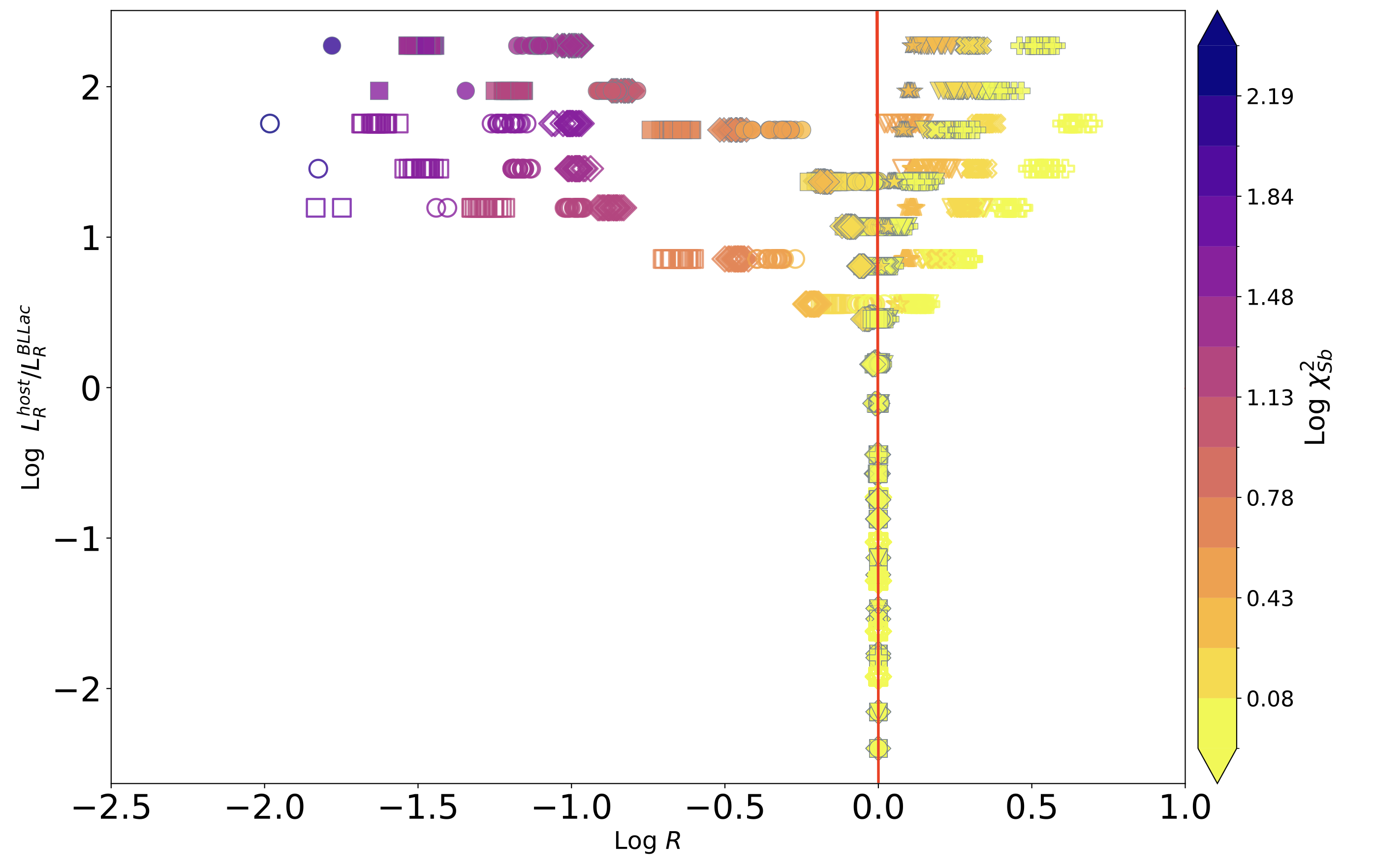}
    \includegraphics[width=0.535\columnwidth]{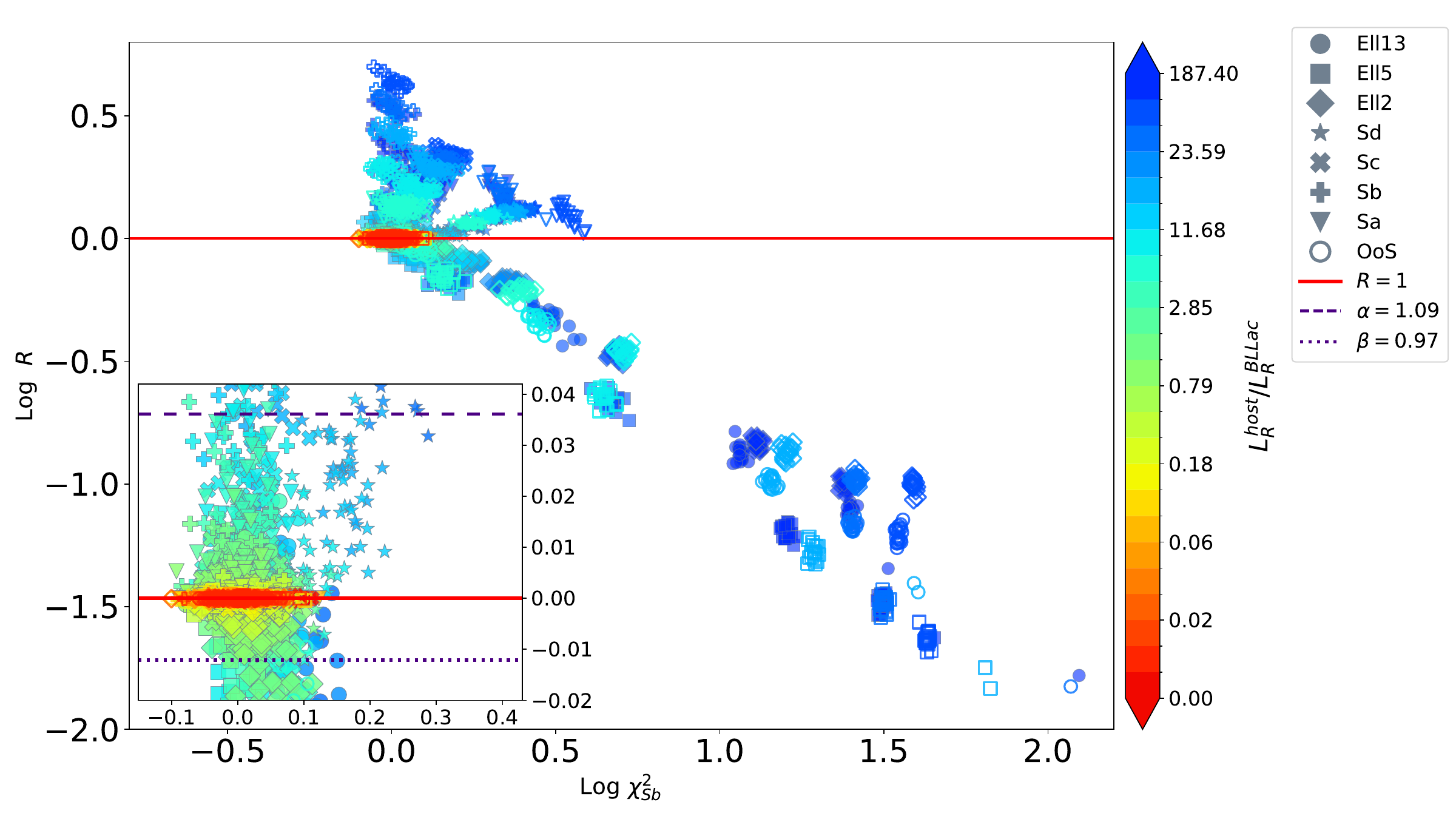}

    \caption{\textit{Left panel:} Distribution of $R$ with respect to the host-to-jet luminosity ratio $L_R^{host}/L_R^{BLLac}$ for the total sample, including the spectra following \citet[][filled symbols]{ghisellini17} and the out-of-sequence spectra (OoS, empty symbols). The color coding shows the distribution of $\chi^2_{Sb}$. The points representing elliptical galaxies are more widely spread away from the $R=1$ line than those representing spiral galaxies, emphasising how easier it is to correctly distinguish an elliptical host than a spiral one. \textit{Right panel:} Ratio of the statistics $R$ as a function of $\chi^2_{Sb}$ for the total sample (including out-of-sequence spectra). The color coding follows Figure \ref{fig:plot_R}.
    }
    \label{fig:R_withshiftedSEDs}
\end{figure}

We performed our analysis on this set following the approach detailed in Section \ref{section:3}, and populated the $R$ vs $\chi^2_{Sb}$ plot. 
Figure \ref{fig:R_withshiftedSEDs} (right panel) shows the results of the total sample, including both results from the sequence and out-of-sequence jet SEDs. 

The upper and lower limits of the confusion region are consistent with what found for the core sample only, i.e.\ $\alpha=1.09$ and $\beta=0.97$.
The confusion region is thus driven by the core sample. 
64.2\% of all sources fall within the confusion region, suggesting that out-of-sequence sources might allow an easier identification of their host galaxies.

As expected, the out-of-sequence spectra populate new parts of the classification plot. 
Most notably, the spectra generated with $S_a$ templates and significant dominance of the host galaxies over the jet emission rightly fall in the ``spiral" part of the plot, but they differentiate from the other spirals, by moving farther away from $\chi^2_{Sb}=1$. 
This is also clearly shown in the left panel of Figure \ref{fig:R_withshiftedSEDs}.
This further shows that when a host galaxy is dominant over the non-thermal continuum, even this simple approach can give not only indications on the general nature of the host, but also their sub-types, at least whether they are compatible or not with the reference model.

\bibliography{BIB}{}
\bibliographystyle{aasjournalv7}

\end{document}